\documentclass[12pt,aps,eqsecnum,amsfonts,amsmath]{article}
\usepackage{amssymb} 
\begin{document}
\newcommand{\PBm}{\Phi_{\rm B}^m} \newcommand{\PH}{\Phi_{\rm H}}
\newcommand{\PA}{\Phi_{\underline\alpha}} \newcommand{\eins}{{\mathbf 1}}
\newcommand{\be}{\begin{equation}} \newcommand{\ee}{\end{equation}}
\newcommand{\PP}{{\cal P}} \newcommand{\QQ}{{\cal Q}}
\newcommand{\NN}{{\mathbb N}} \newcommand{\RR}{{\mathbb R}}
\renewcommand{\v}{\vert} \newcommand{\vv}{\v\!\v}
\newcommand{\CN}{(c_{2n})_{n\in\NN}}
\newcommand{\QED}{\hspace*{\fill}Q.E.D.\vskip2mm}
\renewcommand{\today}{} 
\title{\vskip-15mm \bf On a Class of Bounded Quantum Fields}
\author{{\sc M. Grott}\thanks{Electronic address: 
{\tt grott@theorie.physik.uni-goettingen.de}} \\ and \\
{\sc K.-H. Rehren}\thanks{Electronic address: 
{\tt rehren@theorie.physik.uni-goettingen.de}}
\\[3mm] Institut f\"ur  Theoretische Physik, Universit\"at
  G\"ottingen,\\ 37073 G\"ottingen, Germany}
\maketitle

\begin{abstract}
Local quantum fields in 1+1 dimensions can have bounded field operators. 
The class of such fields which in addition obey Huygens' principle
(time-like commutativity) and conformal covariance, is completely
determined. The result confirms and qualifies a conjecture by K. Baumann.  
\\[1mm] {PACS 03.70.+k,11.10.Cd}
\end{abstract}

\section{Introduction}
Free Bose fields always have unbounded smeared field operators. 
Contrary to naive extrapolation from this fact, Bose fields which have
bounded field operators are known to exist at least in 1+1
dimensions. The simplest examples have been given by Buchholz
\cite{Bu} and are of the form  
$$\PBm(t,x)=\psi_m(t+x)\otimes\psi_m(t-x)
\eqno(1.1)$$
where $\psi_m=\partial^m\psi_0$ is a (real) chiral free Fermi field of 
half-integer scaling dimension $m+\frac12$ (i.e., a derivative of the 
canonical field $\psi_0$ with scaling dimension $\frac12$). These
fields have the additional property that they commute not only at
spacelike distance as required by locality, but also at timelike
distance. We call such fields in the sequel ``Huygens fields'' because
they obey Huygens' principle of propagation along the
light-ray. Further examples of bounded Bose fields {\em not}
obeying Huygens' principle have been given in \cite{KHR}.  

The existence of bounded Bose fields has an important bearing \cite{Y}
on the structure of the Borchers algebra divided by its locality
ideal. This quotient may be considered as the universal algebraic
structure underlying any local quantum field theory \cite{Bor}. On
abstract grounds, it can be equipped with an abundance of operator
topologies. The existence of bounded Bose fields then ensures that
among these topologies there are some which admit a vacuum state. A
vacuum state in this setting is a state which annihilates the spectrum
ideal (Doplicher ideal), that is, the left ideal generated by all
operators with momentum transfer outside the forward light-cone
(annihilation operators). 

The last paper of our late collegue K. Baumann \cite{Bau} was devoted
to the study of conformally invariant scalar bounded Huygens
fields. He obtained the result that for odd scaling dimensions $2m+1$
their truncated $2n$-point functions $W^T_{2n}$ are necessarily
multiples of the truncated $2n$-point functions $V^T_{2n}$ of the
field $\PBm$, with a positive coefficient $c_{2n}$ for every
$n$. (The odd functions vanish identically.) It was clear to Baumann
that Wightman positivity puts severe further constraints on the
sequence of coefficients, and he conjectured that the only solutions
to these constraints are the weighted ``s-products'' \cite{Bor} of the
elementary field $\PBm$. Recall that the s-product of two or more
Wightman fields $\Phi_i$ is another Wightman field, which is defined
equivalently as the sum of the fields $\Phi_i$ acting on the tensor
product Hilbert space $\bigotimes_i H_i$, or by multiplying the vacuum
expectation values $(\Omega,\exp \Phi_i(tF)\Omega)$, or by adding the
truncated Wightman functions of the fields $\Phi_i$.  

Thus, weighted s-products of $\PBm$ are operator valued distributions
of the form  
$$\PA^m(F)=\sum_i \left(\eins^{\otimes(i-1)}
\otimes\PBm(\alpha_i F)\otimes\eins^{\otimes(I-i)}\right)
\eqno(1.2)$$
with $I\in\mathbf N$ or possibly $I=\infty$, and their truncated
Wightman functions are 
$$ W^T_{2n} = \left(\sum_i\alpha_i^{2n}\right) \cdot V^T_{2n}, \qquad
\hbox{i.e.,} \qquad c_{2n}=\sum_i\alpha_i^{2n}.
\eqno(1.3)$$
For every real suitably normalized testfunction $F$ of the form
$F(t,x)=g(t+x)f(t-x)$ the field operator
$\PBm(F)=\psi_m(g)\otimes\psi_m(f)$ is an involution, 
$\PBm(F)^2 = \eins$. For such testfunctions the vacuum expectation
functionals are $(\Omega,\exp \PBm(tF)\Omega) =\cosh t$ and   
$$ (\Omega,\exp \PA^m(tF)\Omega) = \prod_i \cosh\alpha_i t. 
\eqno(1.4)$$

These s-product fields are examples for bounded Huygens fields provided 
$\sum_i\vert\alpha_i\vert<\infty$. 

The resolution of the positivity constraints on the coefficients
$\CN$ for general bounded Huygens fields turned out to be a most
intricate problem, which Baumann was not meant to settle. We shall
solve it in this article.  

Our strategy is the following. In the first step we show that the sequence
$\CN$ defining the given bounded Huygens field, satisfies (i) a
certain growth condition (``exponential boundedness'', Prop.\ 3 below)
as a consequence of the boundedness of field operators, and (ii) an
infinite system of positivity conditions (``determinant positivity'',   
Prop.\ 4) as a consequence of Wightman positivity.

The second step, the actual s-product decomposition, is completely
independent of quantum field theory. We show that every sequence of
numbers $\CN$ satisfying (i) and (ii) is either trivial, $c_{2n}\equiv 0$, 
or there exists a finite or infinite sequence of weights $\alpha_i > 0$ 
such that $c_{2n}=\sum_i\alpha_i^{2n}$ (Prop.\ 8). This is the
solution to a kind of momentum problem. It is constructive to the
extent that the weights $\alpha_i$ can be computed from the sequence $\CN$.    

Since the knowledge of the coefficients $\CN$ determines all Wightman
functions and hence the Wightman field up to unitary equivalence, the
conclusion is that the given Huygens field is indeed a finite or infinite 
weighted s-product of the elementary Huygens field (Thm.\ 9). This
proves Baumann's conjecture. In particular, we implicitly rule out the
speculative existence of ``continuous s-products'' of the fields
$\PBm$ (which exist for free fields \cite{Gui}) since these would 
necessarily violate Wightman positivity for mixed correlations.   

\section{Baumann's Theorem}
We first quote Baumann's result \cite{Bau}, and derive a direct
consequence from it.

\vskip2mm\noindent {\sl {\bf 1. Theorem:} Let $\PH$ be a hermitean
  scalar Bose field in 1+1 spacetime dimensions which is conformally
  covariant with odd integer scaling dimension $d=2m+1$, and which
  commutes with itself at spacelike and at timelike distance. If
  $\PH(F)$ is a bounded operator for every testfunction $F$, then the
  odd Wightman functions of $\PH$ vanish and there exists a sequence
  of positive numbers $\CN$ such that the even truncated Wightman
  functions $W^T_{2n}$ are multiples   
$$ W^T_{2n} = c_{2n} V^T_{2n} \eqno(2.1)$$
  of the truncated Wightman functions $V^T_{2n}$ of the field $\PBm$. }
  \vskip2mm

Examples of fields satisfying the assumptions are the weighted 
s-products $\PA^m$ for which $c_{2n}=\sum_{i} \alpha_i^{2n}$.

The smeared Wightman functions $W_{2N}(F_1,\dots,F_{2N})$ of the field
$\PH$ can be expanded in products of the truncated Wightman functions
$W^T_{2n}$. Using the basic relation (2.1), and expanding back the 
truncated functions $V^T_{2n}$ into the Wightman function $V_{2\nu}$ of 
the elementary field $\PBm$, one obtains a partition expansion of
$W_{2N}$ in terms of products of $V_{2\nu}$ with coefficients which
are certain universal polynomials in the coefficients $\CN$. 

We shall specify this statement in more detail and more in generality,
replacing the individual testfunctions $F_i$ by multiple commutators
of testfunctions in the Borchers algebra (that is, considering test
functions as insertions into products of fields \cite{Bor}). We 
abbreviate a commutator $[\Phi(F_1),\Phi(F_2)]$ by $\Phi([F_1,F_2])$
and extend this notation to arbitrary multiple commutators of
length $l$ such as  
$$K=[\cdots[F_1,F_2],\cdots,F_l]$$
(or any other ordering of the bracketing). A testfunction $F$ is just a
multiple commutator of length $l=1$. Putting 
$$\overline{K}=[\overline{F_l},\cdots,[\overline{F_2},\overline{F_1}]\cdots]$$
with the reverse ordering of the brackets (and $\overline F$ the complex 
conjugate testfunction), we have for a hermitean field 
$$\Phi(K)^* = \Phi(\overline{K}).$$

\vskip2mm\noindent {\sl {\bf 2. Proposition:} Let $\PH$ be a bounded
  Huygens field as in Theorem 1, and let $K_1,\dots,K_N$ be multiple
  commutators of length $l_i$ as above, $\sum_il_i=L$. Then 
$$ W_L(K_1,\dots,K_N) = \sum_{\PP(\NN_N)} c_{\vv P_1\vv,\dots,\vv P_r\vv}
\prod_{P \in\PP} V(P).
\eqno(2.2)$$  
  (Notation: The sum extends over all partitions $\PP$ of the set
  $\NN_N\equiv\{1,\dots,N\}$ into $r$ mutually disjoint nonempty
  subsets $P$, $1\leq r \leq N$. $V(P)$ stands for 
  $V_{\vv P\vv}(K_i,\dots,K_j)$ if $P=\{i,\dots,j\} \subset \NN_N$, 
  $i<\dots <j$, where $\vv P\vv=\sum_{i\in P}l_i$ is the total
  length of all multiple commutators contributing to $V(P)$.) 

  The expansion coefficients $c_{\vv P_1\vv,\dots,\vv P_r\vv}$ are
  universal polynomials in the variables $\CN$, which vanish unless
  all $\vv P_s\vv$ are even. The nonvanishing polynomials    
  $c_{2k_1,\dots,2k_r}$ are determined by the recursion (in $r$)
$$ c_{2k_1,\dots,2k_r,2k_{r+1}}= \sum_{S\subset\NN_r} 
(-1)^{\v S\v}\v S\v!\; c_{2k_{r+1}+\vv S\vv}\cdot c_{2k_t,\dots,2k_u}
\eqno(2.3)$$ 
  where $\vv S\vv=2\sum_{s\in S}k_s$ and
  $\{t,\dots,u\}=\NN_r\setminus S$, and by convention
  $c_\emptyset=1$. They are symmetric in the indices $2k_s$. In
  particular, for $r=1$, the polynomial $c_{2k}$ 
  coincides with the variable $c_{2k}$.} \vskip2mm

{\it Example.} Let $K,L,M,N$ be four commutators of length $3,1,4,6$,
respectively. Then 
\begin{eqnarray*} \lefteqn{W_{14}(K,L,M,N) =  c_{14}\cdot V_{14}(K,L,M,N)
+ c_{8,6}\cdot V_8(K,L,M)V_6(N)+{}} \\ & {}+ 
c_{10,4}\cdot (V_{10}(K,L,N)V_4(M)+V_4(K,L)V_{10}(M,N))
+c_{6,4,4}\cdot V_4(K,L)V_4(M)V_6(N)
\end{eqnarray*}
with $c_{10,4}=c_{10}c_4-c_{14}$, $c_{8,6}=c_{8}c_6-c_{14}$, and
$c_{6,4,4}=c_6c_{4,4}-2c_{10}c_4+2c_{14}=c_6c_4^2-c_8c_6-2c_{10}c_4+2c_{14}$.

We shall refer to the universal polynomials $c_{2k_1,\dots,2k_r}$ in
the variables $\CN$ as ``partition polynomials''. 

\vskip2mm {\it Proof.} First we note that the general expansion
formula for Wightman functions in terms of truncated Wightman
functions extends to multiple commutators:
$$ W_L(K_1,\dots,K_N) = \sum_{\PP(\NN_N)}\prod_{P \in\PP} W^T(P)
\eqno(2.4)$$
with notations as in the proposition, 
$W^T(P) = W^T_{\vv P\vv}(K_i,\dots,K_j)$ if 
$P=\{i,\dots,j\}\subset \NN_N$, $i<\dots <j$, and $L=\sum_il_i$. This
is true because in the usual expansion in terms of individual
testfunctions all contributions with two functions belonging to some
commutator being distributed over different factors $W^T(P)$ cancel
due to antisymmetrization. It follows that the inverse formula also
generalizes to multiple commutators of functions:
$$ W^T_L(K_1,\dots,K_N) = \sum_{\PP(\NN_N)}
(-1)^{r-1}(r-1)!\;\prod_{P \in\PP} W(P) \qquad(r=\v\PP\v).
\eqno(2.5)$$
The expansions (2.4) and (2.5) hold for general Wightman fields, and
hence remain valid if $W$ are replaced by $V$.

Inserting the basic relation (2.1) into the expansion (2.4), and using
the inverse expansion (2.5) to expand $V^T$ in terms of $V$, already
shows that an expansion of the form 
$$ W_L(K_1,\dots,K_N) = \sum_{\PP(\NN_N)} C(\PP) \prod_{P \in\PP} V(P)
\eqno(2.6)$$
holds, with coefficients $C(\PP)$ to be determined as follows. 
Consider the identity
$$ W_{L+l}(K_1,\dots,K_N,K) = \sum_{M\subset \NN_N}W^T(M,K)W(\NN_N\setminus M)
\eqno(2.7)$$
where $W(M,K)=W_{\vv M\vv+l}(K_i,\dots,K_j,K)$ if $M=\{i,\dots,j\}$,
$i<\dots<j$. This identity follows from (2.4) by collecting all
factors $W^T(P)$ which involve the distinguished multiple commutator
$K$ of length $l$. We expand both sides of (2.7) in terms of the Wightman
functions $V$ of the elementary field, by using the expansion (2.6)
($W \to V$), the basic relation (2.1) ($W^T\to V^T$), and the inverse
expansion (2.5) ($V^T\to V$). Equating the resulting coefficients of
products $\prod_P V(P)$ on both sides (with a focus on the
distingushed factor $V(P)$ which contains $K$), we obtain the recursion  
$$ C(\PP\cup\{P\})=\sum_{\QQ\subset\PP}(-1)^{\v\QQ\v}\v\QQ\v!\; 
c_{\vv P\vv+\vv\QQ\vv} C(\PP\setminus\QQ).
\eqno(2.8)$$
Here $\PP$ and $\QQ$ are sets of (mutually disjoint non-empty) subsets
of $\NN_{N+1}\setminus P$, and $C(\emptyset):=1$.

An inspection of the structure of this recursion reveals that $C(\PP)$
depend only on the total lengths $\vv P\vv=\sum_{i\in P}l_i$ of all
$P\in\PP$, that is $C(\PP)=c_{\vv P_1\vv,\dots,\vv P_r\vv}$,
irrespective of their order, and vanish unless all $\vv P_s\vv$ are
even. The recursion (2.3) is then just a transscription of (2.8). The
last statement of the proposition is obvious. \QED 

We note that for an s-product field, the partition polynomials take
the values 
$$c_{2k_1,\dots,2k_r}=
\sum_{i_1,\dots,i_r\hbox{ \small all distinct}}\prod_s\alpha_{i_s}^{2k_s}.
\eqno(2.9)$$

\section{Evaluation of boundedness}

We want to derive a growth condition on the coefficients $\CN$ which
follows from the Huygens field $\PH$ being bounded. For this purpose,
we choose any real testfunction $F$ of the form $F(t,x)=f(t+x)f(t-x)$
as in (1.4). Due to the anticommutation relation, $\psi_m(f)^2$ is a
multiple of $\eins$, so we may normalize $f$ such that $\PBm(F)^2=\eins$, 
and consequently $V_{2n}(F,\dots,F)=1$. 

Thus, the above expansion formula (2.2) for all multiple commutators
equalling $K_i=F$ ($l_i=1$) simplifies to
$$ W_{2n}(F,\dots,F) = \sum_r \sum_{2k_1,\dots,2k_r: \sum k_i=n}
\frac 1{r!} \frac{(2n)!}{(2k_1)!\dots (2k_r)!}\; c_{2k_1,\dots,2k_r} .
\eqno(3.1)$$ 
The combinatorial factors just count the number of partitions arising
with the same coefficient $c_{2k_1,\dots,2k_r}$. Thus, we obtain 
$$ (\Omega,\exp \PH(tF)\Omega) = 
\sum_n \frac{t^{2n}}{(2n)!} W_{2n}(F,\dots,F) 
= \sum_r \frac1{r!} \sum_{2k_1,\dots,2k_r} 
\frac{t^{2k_1+\dots+2k_r}} {(2k_1)!\dots (2k_r)!}\; c_{2k_1,\dots,2k_r}.
\eqno(3.2)$$ 
We denote this power series in $t$ by $E(t)$. Since 
$(\Omega,\exp \PH(tF)\Omega) \leq \exp \v t\v\vv\PH(F)\vv$, we obtain 
the growth condition on the coefficients $\CN$:

\vskip2mm\noindent {\sl {\bf 3. Proposition:} (Exponential boundedness) 
  The power series $E(t)$ defined by the right-hand side of eq.\ (3.2)
  with coefficients which are polynomials in $\CN$, converges for all
  $t$ and is bounded by  
$$ E(t) < \exp \lambda \v t\v
\eqno(3.3)$$ 
  where $\lambda = \vv\PH(F)\vv < \infty$.} 

\section{Evaluation of Wightman positivity}

We want to derive an infinite system of positivity conditions on 
the coefficients $\CN$ which follow from Wightman positivity of the
field $\PH$, i.e., positive definiteness of the Hilbert space inner
product determined by the Wightman functions. For this purpose, we
choose special testfunctions and exploit the anticommutation
relations of $\psi_m$ which imply simple identities for multiple
commutators of $\PBm$, see \cite{Bau}. We fix one complex testfunction
$f_0$ with compact momentum support in $\RR_+$ such that $\psi_m(f_0)$
is a creation operator, and $\psi_m(f_0)^2=0$. We normalize it such
that $\vv\psi_m(f_0)\Omega\vv=2$. 

If $f$ is a real testfunction with momentum support disjoint from
the momentum support of $f_0$, then $\{\psi_m(f),\psi_m(f_0)\}=0$. It
can be normalized such that $2\psi_m(f)^2 = \eins$. We introduce the
multiple commutators of length $k$, 
$$K_k(f)= [\cdots[[(f_0 \times f),(f\times f)],(f\times f)],
\cdots,(f\times f)]$$ 
where $(g\times f)(t,x) = g(t+x)f(t-x)$. An easy induction shows that
the multiple commutators, if evaluated with the field $\PBm$, are
periodic in the length $k$:  
$$\PBm(K_{2n+1}(f))=\psi_m(f_0) \otimes \psi_m(f)
\quad{\rm and}\quad 
\PBm(K_{2n}(f))=\psi_m(f_0) \psi_m(f) \otimes \eins.$$

Now we choose real testfunctions $f_i$ ($i=1,\dots r$) with properties like 
$f$ before ($f_0$ remaining fixed), and with mutually disjoint momentum
supports. Then all products of field operators
$\PBm(K_{k_i}(f_i))\PBm(K_{k_j}(f_j))$ vanish because of $\psi_m(f_0)^2=0$. 
Furthermore, $V(\overline{ K_{k_i}(f_i)},K_{k_j}(f_j))=\delta_{ij}$ 
because $\psi_m(f_0)^*\psi_m(f_0)\Omega = 2\Omega$ and 
$\omega(\psi_m(f_i)\psi_m(f_j))=\frac12 \delta_{ij}$ 
due to the support and normalization assumptions. 

These choices are taylored to the effect that the mixed correlations
of the elementary field $\PBm$, 
$V(\overline{K_{k_{i_m}}(f_{i_m})},\dots,\overline{K_{k_{i_1}}(f_{i_1})},
K_{k_{j_1}}(f_{j_1}),\dots,K_{k_{j_n}}(f_{j_n}))$, vanish unless 
$n=m=1$ and unless $j_1=i_1$. It follows that in the expansion of 
the mixed correlations 
$W(\overline{K_{k_r}(f_r)},\dots,\overline{K_{k_1}(f_1)}, 
K_{k_1}(f_1),\dots,K_{k_r}(f_r))$ according to Prop.\ 2, only the single term 
$c_{2k_1,\dots,2k_r}\prod_s V(\overline{K_{k_s}(f_s)},K_{k_s}(f_s))=
c_{2k_1,\dots,2k_r}$ does not vanish. Hence
$$\vv \PH(K_{k_1}(f_1))\cdots\PH(K_{k_r}(f_r))\Omega\vv^2 
= c_{2k_1,\dots,2k_r}.$$
By Wightman positivity, the partition polynomials $c_{2k_1,\dots,2k_r}$ 
and in particular the coefficients $c_{2k}$ themselves must be nonnegative.  

The argument is easily generalized by choosing an $r\times N$ matrix 
of lengths $k_i^\alpha$. We find 
$V(\overline{K_{k^\alpha_i}(f_i)},K_{k^\beta_j}(f_j))=
\delta_{ij}\delta_{k^\alpha_i,k^\beta_i\; {\rm mod}\; 2}$ 
 such that 
$$M_{\alpha\beta}\equiv 
W(\overline{ K_{k_r^\alpha}(f_r)},\dots,\overline{K_{k_1^\alpha}(f_1)},
K_{k_1^\beta}(f_1),\dots,K_{k_r^\beta}(f_r))= 
c_{k_1^\alpha+k_1^\beta,\dots,k_r^\alpha+k_r^\beta}.$$
Since $M_{\alpha\beta}$ is a matrix of inner products 
$(\Phi^\alpha\Omega,\Phi^\beta\Omega)$ of Hilbert space vectors,
Wightman positivity requires $M_{\alpha\beta}$ to be a positive
semidefinite matrix. We conclude  

\vskip2mm\noindent {\sl {\bf 4. Proposition:} (Determinant positivity)
  For every $r\times N$ rectangular matrix with positive integer
  entries $k_i^\alpha$ the $N \times N$ determinant 
$$\det \left( c_{k_1^\alpha+k_1^\beta,\dots,k_r^\alpha+k_r^\beta}
\right)_{\alpha,\beta} \geq 0
\eqno(4.1)$$
  (a polynomial of polynomials, hence a polynomial in $\CN$) 
  is nonnegative.} \vskip2mm

These are the positivity constraints on the coefficients $\CN$. In
particular, for $r=1$, $N=2$ and $(k^\alpha)=(k-1,k+1)$ we get the constraint
$$ c_{2k-2}c_{2k+2}-c_{2k}^2 \geq 0.
\eqno(4.2)$$
Interestingly enough, the argument that follows will only exploit
$N=1$ and $N=2$ determinant positivity (but $r$ arbitrary).

\section{s-Product decomposition}

We have concluded that the coefficients $c_{2n}$ defining the bounded 
Huygens field $\PH$ satisfy the constraints given in Propositions 3 and 4 
(exponential boundedness and determinant positivity). These two
properties will be the only input throughout this section, while
contact with quantum fields will be only made in the end (Thm.\ 9).
We want to derive the decomposition 
$$c_{2n}=\sum_i \alpha_i^{2n}$$
for any sequence $\CN$ satisfying exponential boundedness and
determinant positivity. We do so by successively extracting the
leading weights $\alpha_i$ in decreasing order, and correspondingly
reducing the given sequence $\CN$. These procedures are guided by the
expected formula (2.9).

First, we need
\vskip2mm\noindent {\sl {\bf 5. Lemma:} If a sequence of numbers
  $\CN$ satisfies determinant positivity and exponential boundedness,
  and if $c_{2n}$ vanishes for some $n$, then it vanishes for all
  $n$.} \vskip2mm 

{\it Proof.} Eq.\ (4.2) implies that $c_{2k}$ vanishes
whenever $c_{2k-2}$ vanishes, or whenever $c_{2k+2}$ vanishes and
$k>1$. But, assuming $c_{2n}=c_2\delta_{n1}$ makes all partition
polynomials $c_{2k_1,\dots,2k_r}$ vanish except $c_{2,\dots,2}=c_2^r$. 
Inserting into the power series $E(t)$, eq.\ (3.2), yields 
$E(t)=\exp\frac{c_2}2t^2$ which violates the exponential bound (3.3)
unless $c_2=0$. \QED

The lemma just confirms the well known fact that if the truncated functions
$W^T_{n}$ of any Wightman field vanish identically for some $n>2$,
then they vanish for all $n>2$, and the field is a generalized free
field. A nontrivial generalized free field is unbounded, so also $W_2$
must vanish for a bounded field.  

If any and hence all $c_{2n}$ are strictly positive, the sequence 
$\left(\frac{c_{2n+2}}{c_{2n}}\right)_{n\in\NN}$ increases mono\-tonously 
due to determinant positivity, eq.\ (4.2). It is also bounded due to 
exponential boundedness. This can be seen as follows. Monotony implies 
$c_{2k} \geq c_{2n}\left(\frac{c_{2n+2}}{c_{2n}}\right)^{k-n}$. Now
discard from the power series $E(t)$ all terms with $r>1$. Since every
coefficient in $E(t)$ is nonnegative, the remaining power series
$E'(t)$ is still exponentially bounded. Hence for any fixed $n$,
$$\cosh \left(\frac{c_{2n+2}}{c_{2n}}\right)^{\frac12}t = 
\sum_k \frac{t^{2k}}{(2k)!}\left(\frac{c_{2n+2}}{c_{2n}}\right)^k 
\leq C \sum_k \frac{t^{2k}}{(2k)!}\; c_{2k} = C E'(t) \leq C E(t) \leq
C \exp\lambda \v t\v$$ 
with a constant $C$ depending on $n$. This implies
$\frac{c_{2n+2}}{c_{2n}} \leq\lambda^2$. Hence the limit 
of the monotonously increasing bounded sequence exists, and   
$$ \alpha := \left(\lim_n \frac{c_{2n+2}}{c_{2n}}\right)^{\frac12}
\leq\lambda.
\eqno(5.1)$$

\vskip2mm\noindent {\sl {\bf 6. Proposition:} The reduced sequence
  $(c^*_{2n})_{n\in\NN}$ with 
$$ c^*_{2n} = c_{2n} - \alpha^{2n}
\eqno(5.2)$$ 
  is again exponentially bounded with the reduced bound 
  $\lambda^* = \lambda-\alpha\geq 0$, and it again satisfies
  determinant positivity.} \vskip2mm 

Before we prove the proposition, let us discuss its consequences. 
The reduction prescription $*$ is an operation on nonvanishing
sequences $\CN$ satisfying exponential boundedness and determinant
positivity, and yields a weight $\alpha$ and another sequence
$(c^*_{2n})_{n\in\NN}$ satisfying exponential boundedness and
determinant positivity. It will became clear in the course of the
proof that it can be interpreted as an ``s-division''
by the leading factor of a weighted s-product, in agreement with the
subtraction formula (5.2). 

Now, if $(c^*_{2n})_{n\in\NN}$ vanishes, we have  
$c_{2n}=\alpha^{2n}$; otherwise the operation $*$ can be iterated.
(Recall that according to Lemma 5, $(c^*_{2n})_{n\in\NN}$ is strictly 
positive if and only if any single $c^*_{2n}\neq 0$.) 
This yields a succession of sequences $((c^{(i)}_{2n})_{n\in\NN})_i$
where $i$ stands for the order of the iteration, starting with
$(c^{(0)}_{2n})_{n\in\NN} = \CN$, a succession of weights $\alpha_i$
starting with $\alpha_1=\alpha$, and a succession of exponential
bounds $\lambda^{(i)}$ starting with $\lambda^{(0)}=\lambda$: 
$$ c^{(I)}_{2n}=c_{2n}-\sum_{i=1}^I\alpha_i^{2n} \qquad\hbox{and}\qquad
\lambda^{(I)}=\lambda-\sum_{i=1}^I\alpha_i\geq 0 .
\eqno(5.3)$$

The iteration stops if $(c^{(I)}_{2n})_{n\in\NN}$ vanishes for some
$I$. Hence $c_{2n}=\sum_{i=1}^I \alpha_i^{2n}$, and we have
established eq.\ (1.3) with a finite sum, i.e., the bounded Huygens
field $\PH$ providing the initial sequence $\CN$ is indeed an s-product.  

If the iteration never stops, we shall show that for each $n$, the numbers 
$c^{(i)}_{2n}$ converge to $0$ as $i$ grows. It follows from (5.3) that 
$c_{2n}=\sum_{i=1}^\infty \alpha_i^{2n}$. In this case, we have 
established that $\PH$ is an infinite s-product. 

{\it Proof of Prop.\ 6.} We consider the limits
$$ c^*_{2k_1,\dots,2k_r} = \lim_n \frac{c_{2k_1,\dots,2k_r,2n}}{c_{2n}}. 
\eqno(5.4)$$
Inserting the recursion relation (2.3), and using
$\lim_n\frac{c_{2n+2k}}{c_{2n}}=\alpha^{2k}$ for any fixed $k$, these
limits exist, and 
$$ c^*_{2k_1,\dots,2k_r} = \sum_{S\subset\NN_r} 
(-1)^{\v S\v}\v S\v!\; \alpha^{\vv S\vv}\cdot c_{2k_t,\dots,2k_u}
\eqno(5.5)$$ 
where as before, $\vv S\vv=2\sum_{s\in S}k_s$ and 
$\{t,\dots,u\}=\NN_r\setminus S$, and $c^*_\emptyset=1$. 

In particular, $c^*_{2n}=c_{2n}-\alpha^{2n}$ in agreement with eq.\
(5.2), but in order to justify the notation for the other limits 
$c^*_{2k_1,\dots,2k_r}$, we have to show that they are indeed the correct
universal polynomials in the variables $(c^*_{2n})_{n\in\NN}$. 
To this end, it is sufficient to show that the limits
$c^*_{2k_1,\dots,2k_r}$ satisfy the recursion relations (2.3) which
uniquely fix them in terms of $(c^*_{2n})_{n\in\NN}$: 
$$c^*_{2k_1,\dots,2k_r,2k_{r+1}} = \sum_{S\subset\NN_r}
(-1)^{\v S\v}\v S\v!\; c^*_{2k_{r+1}+\vv S\vv}\cdot
c^*_{2k_i,\dots,2k_j}.
\eqno(\hbox{2.3*})$$ 

To establish (2.3*), it is convenient to write $c(S)=c_{2k_s,\dots,2k_t}$
for a subset $S=\{s,\dots,t\}$ of $\NN_{r+1}$, and likewise for
$c^*$. Let $k=k_{r+1}$.   

On the left-hand side of (2.3*), we split the expansion of
$c^*(\NN_{r+1})$ according to eq.\ (5.5) into the sum over those
subsets $T$ of $\NN_{r+1}$ which do not contain the element $r+1$, and
the sum over those subsets $T'=T\cup\{r+1\}$ which do contain
$r+1$. On the right-hand side, we separate the sum into a difference
of two sums corresponding to the two contributions to 
$c^*_{2k+\vv S\vv}= c_{2k+\vv S\vv}-\alpha^{2k+\vv S\vv}$. Then we  
show separate equality of the respective terms,
$$\sum_{T\subset\NN_r} (-1)^{\v T\v}\v T\v!\;
\alpha^{\vv T\vv}\cdot c(\NN_{r+1}\setminus T) = 
\sum_{S\subset\NN_r} (-1)^{\v S\v}\v S\v!\; 
c_{2k+\vv S\vv}\cdot c^*(\NN_r\setminus S)$$ 
and 
$$\sum_{T\subset\NN_r} (-1)^{\v T\v+1}(\v T\v+1)!\;
\alpha^{2k+\vv T\vv}\cdot c(\NN_r\setminus T)
= -\sum_{S\subset\NN_r} (-1)^{\v S\v}\v S\v!\;
\alpha^{2k+\vv S\vv}\cdot c^*(\NN_r\setminus S). 
$$
Inserting (5.5) into the right-hand sides of these equations and
rearranging the summations over subsets, the first one reduces to the
recursion relation (2.3) which holds for $c_{2k_1,\dots,2k_r}$, while
the second one reduces to the combinatorial identity for the subsets of a 
set $T$, $\sum_{U\subset T}\v U \v!\;\v(T\setminus U)\v! = (\v T\v+1)!$.
 
Thus the limits $c^*_{2k_1,\dots,2k_r}$ again satisfy the recursion
relation (2.3*), and consequently may be safely regarded as the partition
polynomials in the variables $(c^*_{2n})_{n\in\NN}$.  

Now, the approximating ratios in (5.4) for each finite $n$ satisfy
determinant positivity as in (4.1), since this is just another
instance of determinant positivity for $c_{2k_1,\dots,2k_{r+1}}$ with
index chains of length $r+1$. It follows that the limits
$c^*_{2k_1,\dots,2k_r}$ also satisfy determinant positivity.

It remains to prove their exponential boundedness with
$\lambda^*=\lambda-\alpha$ (we already know that $\alpha\leq\lambda$
hence $\lambda^*\geq 0$). Consider the reduced power series $E^*(t)$
obtained by replacing $c_{2k_1,\dots,2k_r}$ by $c^*_{2k_1,\dots,2k_r}$
in (3.2). The claim follows from the identity
$$ E(t)=\cosh \alpha t \cdot E^*(t).
\eqno(5.6)$$
This identity in turn is obtained by inserting the inversion of (5.5),
$$ c_{2k_1,\dots,2k_r} = c^*_{2k_1,\dots,2k_r} +  \sum_{i=1}^r 
\alpha^{2k_i}\cdot c^*_{2k_1,\dots,2k_{i-1},2k_{i+1}\dots,2k_r},
\eqno(5.7)$$ 
into the power series $E(t)$, rearranging the summations, and using
$1+\sum_{k=1}^\infty\frac{(\alpha t)^{2k}}{(2k)!}=\cosh \alpha t$. 
\QED 

We note that the factor $\cosh \alpha t$ in eq.\ (5.6) equals the
vacuum expectation value $(\Omega,\exp \alpha \PBm(tF)\Omega)$ of the
field $\PBm$, contributing to an s-product with weight $\alpha$, cf.\
eq.\ (1.4). 

As discussed above, we are left with the alternative that either
$(c^*_{2n})_{n\in\NN}$ is trivial, or the reduction can be iterated,
giving rise to a succession of sequences $((c^{(i)}_{2n})_{n\in\NN})_i$
and a sequence of weights $(\alpha_i)_i$. In the latter case, we have

\vskip2mm\noindent {\sl {\bf 7. Proposition:} As long as the iteration
  goes, one has  
$$ \alpha_i\leq\alpha_{i-1} \qquad \hbox{and}\qquad
\sum_i\alpha_i\leq\lambda, 
\eqno(5.8)$$
  i.e., the succession of weights is decreasing and absolutely
  summable. If the iteration never stops, then
$$ \lim_i \alpha_i = 0\qquad\hbox{and}\qquad \lim_i c^{(i)}_{2n}=0
  \quad\hbox{for each $n\in\NN$}.
\eqno(5.9)$$
}\vskip2mm

{\it Proof.} Inserting (5.2) into $c_{2n+2}\leq\alpha^2c_{2n}$ gives
$c^*_{2n+2}+\alpha^{2n+2}\leq\alpha^2(c^*_{2n}+\alpha^{2n})$. Thus, the 
ratios $\frac{c^*_{2n+2}}{c^*_{2n}}$ are bounded by $\alpha^2$. It
follows that the next weight in the succession is bounded by the
previous one, proving the first of eq.\ (5.8). The second of eq.\
(5.8) follows from eq.\ (5.3). Eq.\ (5.3) also implies that for every
fixed $n$, the sequence $(c^{(i)}_{2n})_{i}$ decreases monotonously
with $i$ while always remaining positive, hence it converges. For any
fixed $I$ it follows 
$$\lim_i c^{(i)}_{2n+2}\;\leq\; c^{(I)}_{2n+2}\;\leq\; 
\alpha_{I+1}^2 c^{(I)}_{2n}\;\leq\; \alpha_{I+1}^2c_{2n},$$
and since $\alpha_I$ converge to zero, we conclude $\lim_i c^{(i)}_{2n}=0$ 
except possibly $\lim_i c^{(i)}_{2}=:\gamma\neq 0$. As in the argument
leading to Lemma 5, this implies that the corresponding power series
$E^{(\infty)}$ equals $\exp(\frac\gamma2 t^2)$. But this contradicts 
the exponential bound
$$\prod_{i=1}^{I}\cosh (\alpha_it) E^{(\infty)}(t)\leq
\prod_{i=1}^{\infty}\cosh (\alpha_it)
E^{(\infty)}(t)=E(t)\leq\exp\lambda \v t\v. $$
Hence also $\lim_i c^{(i)}_{2}= 0$, completing the proof. \QED

As corollaries, we obtain the desired proposition on sequences
$\CN$, as well as the announced theorem on bounded Huygens fields. 

\vskip2mm\noindent {\sl {\bf 8. Proposition:} For every sequence
  $\CN$ satisfying determinant positivity and exponential
  boundedness with $\lambda<\infty$ there is a sequence of 
  positive weights $(\alpha_i)_i$ such that 
$$ c_{2n}=\sum_i \alpha_i^{2n}. 
\eqno(5.10)$$
  The sum is finite or infinite depending on whether the iterated
  reduction prescription stops after a finite number of iterations, or
  not, and $\sum_i\alpha_i\leq \lambda$. (The weights $\alpha_i$ are
  obtained by iteration of (5.1) and (5.4).)} \vskip2mm

The converse statement is also true, but not very exciting in view of
the explicit formulae (2.9) and $E(t)=\prod_i\cosh\alpha_it$ which
follow from eq.\ (5.10) and directly imply exponential boundedness and
determinant positivity.  

\vskip2mm\noindent {\sl {\bf 9. Theorem:} Every hermitean scalar
  bounded Huygens field $\PH$ of odd scaling dimension $2m+1$ is a
  (possibly infinite) s-product of the elementary Huygens field
  $\PBm$ (cf.\ (1.1)) of the same scaling dimension.} 

\section{Comments}
We have shown that a hermitean scalar bounded Huygens field of odd
scaling dimension is an s-product. The assumption of $\PH$ being
hermitean can be easily dropped, since both the real and imaginary
parts of a bounded Huygens field are bounded Huygens fields. Also the
assumption of being scalar can be relaxed by admitting possibly
different half-integer left and right scaling dimensions
$m_\pm+\frac12$. Baumann's theorem, as well as the rest of the
argument leading to our Theorem 9, will remain true if $\PBm$ is
replaced by the field $\partial_+^{m_+}\partial_-^{m_-}\Phi_{\rm B}^0
=  \psi_{m_+}(t+x)\otimes\psi_{m_-}(t-x)$ of helicity $m_+-m_-$. Thus,
every bounded Huygens fields with half-integer left and right scaling
dimensions is an s-product. It is not known whether such fields exist
with integer left and right scaling dimensions. 

This work is based on the Diploma thesis of the first author \cite{Gro}.


\small
\begin{thebibliography}{9} \itemsep-2pt
\bibitem{Bu} D. Buchholz: unpublished.
\bibitem{KHR} K.-H. Rehren: {\it Bounded Bose fields}, Lett.\ Math.\
  Phys.\ {\bf 40}, 299-306 (1997).
\bibitem{Y} J. Yngvason: {\it Invariant states on Borchers' tensor
    algebra}, Ann.\ Inst.\ H.\ Poincar\'e {\bf 45}, 117-145 (1986). 
\bibitem{Bor} H.-J. Borchers: {\it Algebraic aspects of Wightman field
    theory}, in: Statistical Mechanics and Field Theory, R. Sen and
  C. Weil (eds.)\ Haifa Lectures 1971. New York, Halsted Press, 1972.
\bibitem{Bau} K. Baumann: {\it Bounded Bose fields in 1+1 dimensions
    commuting for space and time like distances}, J.\ Math.\
  Phys.\ {\bf 40}, 1719-1737 (1999). 
\bibitem{Gui} A. Guichardet: {\it Alg\`ebres d' observables associ\'ees
    aux relations de commutation}, Collection Intersciences. Librairie
  Armand Collin, Paris, 1968. 
\bibitem{Gro} M. Grott: {\it s-Produkt-Zerlegung beschr\"ankter
    Bose-Felder in 1+1 Dimensionen}, Diploma thesis, G\"ottingen, 2000
  (in German). 



\end{thebibliography}
\end{document}